\documentclass{article}

\usepackage[T1]{fontenc}
\usepackage[utf8]{inputenc}
\usepackage{xcolor}
\usepackage{lipsum}
\usepackage{tabularx}
\usepackage{boldline}
\usepackage{epigraph}
\usepackage{xspace}
\usepackage{ltxtable}
\usepackage[colorlinks,linkcolor=blue]{hyperref}
\usepackage{breakurl}
\usepackage[small]{dgruyter}
\usepackage{nopageno}

\makeatletter
\def\@tablefont{\normalfont}
\makeatother

\usepackage{listings}
\newcommand\nbrealrickcoll{23\xspace}

\renewcommand\tablefont{}

\begin{document}

\title{Exhaustive Survey of Rickrolling in Academic Literature}
\author[1]{Benoit Baudry}
\affil{KTH Royal Institute of Technology, \texttt{baudry@kth.se}}
\author[2]{Martin Monperrus}
\affil{KTH Royal Institute of Technology, \texttt{monperrus@kth.se}}
\date{February 2022}

\maketitle

\begin{abstract}
BAbstract: Rickrolling is an Internet  cultural phenomenon born in the mid 2000s. Originally confined to Internet fora, it has spread to other channels and media. In this paper, we hypothesize that rickrolling has reached the formal academic world. We design and conduct a systematic experiment to survey rickrolling in the academic literature. As of March 2022, there are \nbrealrickcoll academic documents intentionally rickrolling the reader. Rickrolling happens in footnotes, code listings, references. We believe that rickrolling in academia proves inspiration and facetiousness, which is healthy for good science. This original research suggests areas of improvement for academic search engines and calls for more investigations about academic pranks and humor.
\end{abstract}
~\\

\noindent \emph{Peer-reviewed, published in the proceedings of SIGBOVIK, 2022.\\ \url{http://www.sigbovik.org/2022/proceedings.pdf}.}

\section{Introduction}

\epigraph{``It looks like there aren't many great matches for your search'', Google Search Message}

Rickrolling consists of hiding facetious links on the Internet pointing to a specific Youtube video: the video clip of ``Never Gonna Give You Up'' by singer Rick Astley. The interested reader can watch the preparatory material at
\url{https://youtu.be/dQw4w9WgXcQ}.

Rickrolling is a massive cultural phenomenon. It has appeared on the Internet  in the mid 2000s reaching all countries with an exponential growth. Rickrolling has contributed to the staggering number of views of Rick Astley's Youtube video, 1.17 billion at the time of writing.

In this survey, we address an open research question: has rickrolling  rippled beyond Internet chatting platforms? In particular, we hypothesize that rickrolling reached some formal circles, including the academic world of peer reviewed articles, formally edited books and seriously examined theses.
To study this important question, we follow a rigorous method, articulated around two steps, in order to survey rickrolling in the academic literature (\autoref{sec:methodology}). 

Our results are clearcut: as of March 11 2022, there are 34 academic documents with a mention to Rick Astley's video. In this paper, we exhaustively study them, one by one, to identify the intent and form of the rickrolling (\autoref{sec:results}).

We find that:
1) rickrolling is present in the academic literature, with \nbrealrickcoll academic documents intentionally rickrolling the reader;
2) rickrolling is mostly done in Master's and PhD thesis;
3) rickrolling is performed in majority in the field of information technology.
Our study also uncovers some essential challenges for engineering academic search engine engines 
(\autoref{sec:discussion}).

\section{Methodology}
\label{sec:methodology}

The survey is performed per the state of the art of literature reviews. In this section, we document our experimental methodology. 

\subsection{Research Questions}

Our survey is articulated around 4 research questions

\begin{itemize}
    \item  RQ1. How many unique academic publications refer to the video clip of Rick Astley's song ``Never Gonna Give You Up''?

    \item  RQ2. How many of these academic publications refer to Rick Astley's video clip with the intention of rickrolling their readers?

    \item  RQ3. What is the nature of the academic publications that rickroll?

    \item  RQ4. How is the rickroll integrated into the publication?
\end{itemize}

\subsection{Data Collection}

The systematic methodology that we rigorously follow is structured in 2 steps. 

First, we identify the canonical rickroll url. For this, we enter one single query in Youtube's search engine: ``rick astley never gonna give you up'' and collect the most viewed URL, which is \url{https://youtu.be/dQw4w9WgXcQ}. 

Second, we make one single query in Google Scholar to query all the documents referenced by this single academic search engine containling the term ``\texttt{dQw4w9WgXcQ}'': \url{scholar.google.com/?q=dQw4w9WgXcQ}. This query has been done on Friday 11 March 2022, at 15h02, from a Swedish IP, in Konstfack, Stockholm. The replication package is available on Github at  \href{https://github.com/monperrus/replication-package-exhaustive-systematic-review-rickrolling}{replication-package-exhaustive-systematic-review-rickrolling}.

\begin{longtable}{p{.5cm}|p{8.1cm}|p{.8cm}|p{.6cm}}
    Ref.     &  Title and Comment & Type & Year \\
    \hline
    \cite{islam2022multi} & \href{https://link.springer.com/content/pdf/10.1007\%2F978-981-16-6636-0.pdf}{Multi-label Emotion Classification of Tweets Using Machine Learning}\newline
    Note: rickrolling; part of a made-up tweet in a table. & Article & 2022\\
    \cite{rozga2018adaptive} & \href{https://link.springer.com/content/pdf/10.1007\%2F978-1-4842-3540-9.pdf}{Practical Bot Development}$^1$ \newline Note: rickrolling; part of a made up JSON code listing. & Book & 2018 \\
    \cite{soto2021improving} & \href{https://kilthub.cmu.edu/articles/thesis/Improving_Patch_Quality_by_Enhancing_Key_Components_of_Automatic_Program_Repair/14546868/files/27912276.pdf}{Improving Patch Quality by Enhancing Key Components of Automatic Program Repair}\newline Note: rickrolling; as a foonote in  introduction. & PhD & 2021\\
    \cite{zetterlund2021harvesting} & \href{https://arxiv.org/pdf/2112.08267}{Harvesting Production GraphQL Queries to Detect Schema Faults}\newline Note: rickrolling; as part of a made-up JSON code listing. & Article & 2022 \\
    
    \cite{chandakaudience} & \href{https://shubhamchandak94.github.io/reports/ee192t_report.pdf}{Audience Feedback Final Report} Note: rickrolling; in a student report, 3 rickrolls, once as footnote and twice in the technical appendix. & Report & 2021  \\
    
    \cite{janssens2021data} & \href{https://www.google.com/books?id=-Bg-EAAAQBAJ}{Data Science at the Command Line}\newline Note: rickrolling; as part of a command line example, with a bitly URL specifically created \url{https://bit.ly/2XBxvwK} & Book & 2021\\
    
    \cite{denniss2016never} & \href{https://thecoalhub.com/wp-content/uploads/attach_217.pdf}{Never gonna dig you up! Modelling the economic impacts of a moratorium on new coal mines}\newline Note: rickrolling; in footnote, with a tribute title. & Report & 2016\\
    \cite{chen2010leet}     & \href{https://evols.library.manoa.hawaii.edu/bitstream/10524/2118/1/chen.dissertation.leet_noobs.pdf.pdf}{Leet Noobs: Expertise and Collaboration in a ``World of Warcraft'' Player Group as Distributed Sociomaterial Practice}\newline Note: legit. & PhD & 2010 \\
    \cite{wang2014lifecycle}     & \href{https://dspace.mit.edu/bitstream/handle/1721.1/97377/910739655-MIT.pdf;sequence=1}{Lifecycle of viral Youtube videos}\newline Note: legit. & MSc & 2014 \\
    \cite{budd2016beautiful}     & \href{https://link.springer.com/content/pdf/10.1007\%2F978-1-4302-5864-3.pdf}{CSS Mastery}$^1$\newline Note: rickrolling; as part of a HTML code listing. & Book & 2016 \\
    \cite{easdale2021good}     & \href{https://ir.library.oregonstate.edu/downloads/kp78gq01t}{Good Internet Would be Pretty Cool: A Policy Proposal to Expand Internet Access}\newline Note: rickrolling; reading ``For a video presentation of this paper, please visit this link'' & BSc & 2020\\
    \cite{lindstrom2017mapping}     & \href{https://www.diva-portal.org/smash/get/diva2:1109739/FULLTEXT01.pdf}{Mapping the current state of SSL/TLS}\newline Note: rickrolling; as part of a data example. & BSc & 2017\\
    \cite{silva2020digital} & \href{https://repositorio.iscte-iul.pt/bitstream/10071/21822/1/Master_Francisco_Matos_Silva.pdf}{Digital platform for psychological assessment supported by sensors and efficiency algorithms}\newline  Note: rickrolling, taken as an example URL. & MSc & 2020\\
    \cite{lerner2015analyzing} & \href{https://yoshikohno.net/papers/qr-mobisys2015.pdf}{Analyzing the use of quick response codes in the wild}\newline Note: rickrolling and legit; appears twice, as a QR code in the intro (rickroll), as an explanation in the result section (legit) & Article & 2015 \\
    \cite{sanchez2016mangarizer} & \href{https://riunet.upv.es/bitstream/handle/10251/59553/S\%C3\%\%81NCHEZ\%20-\%20Mangarizer\%3AAplicaci\%C3\%B3n\%20Android\%20para\%20crear\%20manga\%20a\%20partir\%20de\%20un\%20archivo\%20de\%20v\%C3\%ADdeo.pdf}{Mangarizer: Aplicaci{\'o}n Android para crear manga a partir de un archivo de v{\'\i}deo}\newline Note: legit. & MSc & 2016 \\
    \cite{hagen2021enhancing} & \href{https://journals.sagepub.com/doi/pdf/10.1177/13548565211010481}{Enhancing\# TdF2017: Cross-media controversies and forensic fandom during live sports events}\newline Note: rickrolling; as an example URL, in footnote. & Article & 2021\\
    \cite{van2017techniques} & \href{https://www.os3.nl/_media/2016-2017/courses/rp1/p59_report.pdf}{Techniques for detecting compromised IoT devices}\newline Note: : rickrolling,  hidden as fake data in appendix. & MSc & 2017 \\
    \cite{ong2021hard} & \href{http://hrlr.law.columbia.edu/files/2021/12/5_Ong.pdf}{Hard Drive Heritage: Digital Cultural Property in the Law of Armed Conflict}\newline Note: legit. & Article & 2021\\
    \cite{delbruel2017towards} & \href{https://tel.archives-ouvertes.fr/tel-01523568/file/DELBRUEL_Stephane.pdf}{Towards an architecture for tag-based predictive placement in distributed storage systems}\newline Note: legit. &  PhD & 2017 \\
    \cite{vico2021daljinsko}  & \href{https://zir.nsk.hr/islandora/object/fsb:7078/datastream/PDF/download}{Daljinsko upravljanje i nadzor pneumatskog manipulatora}\newline Note: rickrolling, fake reference as last reference. & MSc & 2021\\
    \cite{mattsson2015exhaust} & \href{https://lup.lub.lu.se/luur/download?fileOId=5367695\&func=downloadFile\&recordOId=5367694}{Exhaust gas recirculation on twin shaft gas turbines}\newline Note: : rickrolling, in a fake reference. & MSc & 2015\\
    
    \cite{linder2019grounded} & \href{https://oaktrust.library.tamu.edu/bitstream/handle/1969.1/185080/LINDER-DISSERTATION-2019.pdf?sequence=1\&isAllowed=y}{Grounded Visual Analytics: A New Approach to Discovering Phenomena in Data at Scale}\newline  Note: rickrolling; as part of an example {URL} in a PhD thesis. &PhD & 2019 \\
    
    \cite{kosakowski2013egzystencjalna} & \href{http://repozytorium.amu.edu.pl:8080/bitstream/10593/8706/1/Micha\%C5\%82_Kosakowski_-_Egzystencjalna_Teoria_Umys\%C5\%82u_Konstrukcja_narz\%C4\%99dzia_pomiarowego_2013_MY\%C5\%9ALENIE_TELEOLOGICZNE_INTENCJONALNO\%C5\%9A\%C4\%86.pdf}{Egzystencjalna teoria umys{\l}u. Konstrukcja narz{\k{e}}dzia pomiarowego}\newline  Note: rickrolling; fake reference. & MSc & 2021 \\
    
    \cite{perales2021musica} & \href{http://repositori.uji.es/xmlui/bitstream/handle/10234/194587/TFM_2021_PeralesRozale\%CC\%81n_Berta.pdf?sequence=1\&isAllowed=y}{M{\'u}sica para la prevenci{\'o}n del acoso escolar (ciberbullying) en educaci{\'o}n secundaria}\newline  Note: legit. & MSc & 2021\\
    
    \cite{mick2019one} & \href{https://www.researchgate.net/profile/Kenneth-Mick-Evans/publication/334052666_One_Does_Not_Simply_Preserve_Internet_Memes_Preserving_Internet_Memes_Via_Participatory_Community_-Based_Approaches/links/5d14b8bc92851cf4404f6231/One-Does-Not-Simply-Preserve-Internet-Memes-Preserving-Internet-Memes-Via-Participatory-Community-Based-Approaches.pdf}{One Does Not Simply Preserve Internet Memes: Preserving Internet Memes Via Participatory Community-Based
Approaches}\newline Note: legit. & BSc & 2021 \\
    \cite{scull2015hephaestus} & \href{https://www.cl.cam.ac.uk/~ms705/projects/dissertations/2015-ams247-hephaestus.pdf}{Hephaestus: a Rust runtime for a distributed operating system}\newline note: rickrolling; in a Rust code listing, in appendix. & BSc & 2015\\
    \cite{mittermeier2020cultural} & \href{https://library.oapen.org/bitstream/handle/20.500.12657/47348/external_content.pdf}{A cultural History of the Disneyland theme parks}\newline Note: rickrolling; as footnote, in a multilevel tribute to Star Trek and Terry Pratchett. & Book & 2020 \\
    \cite{fakeref} & Duplicate of \cite{mattsson2015exhaust}, because of a badly referenced subtitle. & - & - \\
    
    \cite{huxtable2014unbuckle} & \href{https://www.cl.cam.ac.uk/research/srg/netos/camsas/pubs/part2-project-unbuckle-mjh233.pdf}{Unbuckle: Faster in the kernel} \newline Note: rickrolling; in code listing, as part of the Appendix. & BSc & 2014 \\
    
    \cite{fry2011forensic} & \href{https://spectrum.library.concordia.ca/id/eprint/7769/1/Fry_MASc_F2011.pdf}{A Forensic web Log Analysis Tool: Techniques and implementation}\newline Note: rickrolling; in a code listing. & MSc & 2011\\
    
    \cite{loiola2018recomendado} & \href{https://repositorio.ufmg.br/bitstream/1843/BUOS-B6GEZC/1/disserta__o_daniel_loiola__final_.pdf}{Recomendado Para Voc{\^e}: o impacto do algoritmo do YouTube na forma{\c{c}}{\~a}o de bolhas}\newline Note: legit; in Brazilian portuguese. & MSc & 2018\\
    
    \cite{lerner2017measuring} & \href{https://digital.lib.washington.edu/researchworks/bitstream/handle/1773/40010/Lerner_washington_0250E_17519.pdf?sequence=1\&isAllowed=n}{Measuring and Improving Security and Privacy on the Web: Case Studies with QR Codes, Third-Party Tracking, and Archives}\newline Note: legit; overlapping content with \cite{lerner2015analyzing} & PhD & 2015\\

    \cite{magnus2019brettspillbasert} & \href{https://ntnuopen.ntnu.no/ntnu-xmlui/bitstream/handle/11250/2617776/no.ntnu:inspera:2326491.pdf?sequence=1}{Brettspillbasert oppl{\ae}ring i informasjonssikkerhet}\newline Note: rickrolling; the Youtube link is given as part of a test to learn to distinguish {URL}s that look suspicious.  & BSc & 2019\\
    
    \cite{helle2020ungdom} & \href{https://www.duo.uio.no/bitstream/handle/10852/80317/1/Masteroppgave-Linus-C--S--Helle.pdf}{Ungdom i en digital verden--en studie om tid, s{\o}vn og dataspill}\newline Note: rickrolling; fake last reference, never cited in the thesis; in Norwegian.  & MSc & 2020\\
    
    \caption{Exhaustive Data Analysis of the Rickrolling Academic Literature. $^1$ means that the Google Scholar title metadata was wrong, and fixed in this table.}  \label{tab:my_label}  \end{longtable}
    %

\section{Experimental Results}
\label{sec:results}


\autoref{tab:my_label} provides the exhaustive list of academic documents that refer to the Youtube video of ``Never Gonna Give you Up''. The references appear in the order given by Google Scholar, in our research-oriented web browsers.

For each document, the first column gives the bibliographic reference. The second column provides the title of the document, as well as a note about the intention behind the presence of the Youtube url. The intention can either be rickrolling, or a legitimate usage of the url, e.g., in the case of academic studies that analyze Internet's culture.
The third column is the publication type and the fourth the publication year.

\textbf{RQ1 (Prevalence)} Per the data collected on March 11 2022, at 15h02, Google Scholar knows 34 documents containing an explicit mention to the Youtube URL for ``Never Gonna Give You Up''. With manual analysis, we identify that two URLs are duplicate. This results in exactly 33 verified documents containing ``\texttt{dQw4w9WgXcQ}''.

\textbf{RQ2 (Intention)} Among the 33 documents, there are 10 articles for which the appearance of the Youtube link is legit, as part of a discussion about rickrolling or Internet memes. Based on a careful manual assessment, we confirm that there are 24 academic documents for which the intention is clearly to rickroll the reader, with no relationship between the topic of the document and the link. This means that rickrolling is significantly more practiced (33x) than studied (10x) in the academic literature. 

\textbf{RQ3 (Publication type)} 
The references to the Youtube link are essentially present in academic theses. We found a total of 22 theses among the 33 documents surveyed in this work, which are distributed as follows: 5 PhD, 11 MSc, 6 BSc theses. 
The other references appear in 4 books, 2 reports and 5 articles. 
We also note that rickrolling in the academic literature is a rather new phenomenon, with only 8 references before 2017, while 25 references have appeared in the last 5 years. The recent growth of rickrolling in the academic literature, combined with a majority of references in theses, likely reflect a generational movement: the BSc and MSc students from the late 2020' were teenagers at the boom of rickrolling in the end of the 2010's.

\textbf{RQ4 (Rickroll form)} 
Per our manual analysis, we are able to create a taxonomy of rickrolling forms.\\
\emph{Footnote} For writers who don't dare to break academic seriousness in the main text \footnote{For a study of academic pranks, we refer the reader to ``Le rire de la vielle dame.'' (Pierre Verschueren), see \url{https://bit.ly/2XBxvwK}.}, it is natural to use footnotes as a place for rickrolling. For example, footnote 36, page 52 of \cite{mittermeier2020cultural} reads ``The original Star Trek series (1966–69) also proclaims space as the final frontier in its opening credits. See \url{https://www.youtube.com/watch?v=dQw4w9WgXcQ}''. \\
\emph{Code} We have seen a number of rickroling cases planted in code listings. For example, Zetterlund et al. \cite{zetterlund2021harvesting} rickroll in a JSON listing, which we reproduce in \autoref{lst:graphql-response}, with permission.\\
\emph{URL} When one needs an example URL or a metasyntactic link, it is a good opportunity to rickroll.  For example, Hagen takes the innocuous example ``when user ‘@salhagen1’ references the link ‘https://www.youtube.com/dQw4w9WgXcQ’ two times or more, we only kept the most-liked instance.'' \cite{hagen2021enhancing}.\\
\emph{References} Finally, one can hide rickrolling links in a reference \cite{fakeref}. For example, the last reference of Helle's master's thesis is a fake rickrolling reference \cite{helle2020ungdom}.

\begin{lstlisting}[label={lst:graphql-response}, caption={An example of rickrolling hidden in a code listing \cite{zetterlund2021harvesting} (reproduced with permission).}, float, numbers=none]
{
  "data": {
    "teasers": [
      { 
        "title": "Finance 101",
        "subTitle": "The basics of finance",
        "url": "https://youtu.be/dQw4w9WgXcQ",
        "__typename": "Teaser"
      },
      { 
        "title": "Development 101",
        "subTitle": null,
        "url": "https://youtu.be/jNQXAC9IVRw",
        "__typename": "Teaser"
      }
    ]
  }
}
\end{lstlisting}

\section{Discussion}
\label{sec:discussion}

\subsection{Threats to Validity}

Our measurement of rickrolling is sound but conservative. Not all rickrolling instances can be found with the Youtube identifier.

First Rick Astley's video clip was released in 1987 and, \href{https://en.wikipedia.org/wiki/4chan}{4chan}, where rickrolling likely was born, went online in 2003. We may miss early rickrolling that predates Youtube (2005).

Second, there are many copycats and remixes of ``Never Gonna Give You Up'' on Youtube. It is clear that some rickrolling documents refer to another URL than the canonical one. 

Third, some academic documents rickroll  with an implicit hyperlink, where the target of the hyperlink does not appear in text. In that case, Google Scholar does not index the document under the rickrolling identifier, and thus does not return it for the query under consideration in this survey. To our knowledge, there is no publicly available way to overcome this. Only a Googler with internal access to the database could be able to see the actual rickrolling hypergraph.

\subsection{Limitations of Indexing}

We note that not all academic search engines are equal. For the same request with the ``\texttt{dQw4w9WgXcQ}'' Youtube video identifier, the IEEE academic search engine yields no result, and the ACM academic search engine returns a single paper \cite{lerner2015analyzing}. This means that Google Scholar is far better at indexing rare terms.

We speculate that academic indexing at IEEE and ACM is limited to known terms from a predefined dictionary, with some stemming.
This is a real limitation. Beyond rickrolling there are many use cases for search for rare terms. For example, researchers may search for rare protein or asteroid identifiers: this may not be supported by IEEE or ACM. For such use cases, Google Scholar provides a great service to science by indexing arbitrary terms. 

Google Scholar is not perfect though. In our study, we have identified two shortcomings in Google Scholar: 1) properly handling subtitles to deduplicate documents 2) properly identifying whether theses are master's thesis or PhD thesis. This hinders the soundness and completeness of all systematic literature reviews based on Google Scholar. In this paper, we address these shortcomings through a complete, thorough manual check of all documents.

Last but bot least, as stated in the threats to validity, a useful feature of an academic search engine is missing, even in the best-of-breed Google Scholar: let users search for all documents with an hyperlink to a particular URL, or with a term contained in that hyperlink. This information is most certainly available in Google's database but is not exposed in their public API.

\section{Conclusion}
In this paper, we have presented the first ever study of rickrolling in the academic literature.
Although rickrolling in academia remains confidential, it is clearly an inspiring force for students and scholars alike. This is evidenced by a significant growth of rickrolling in the last 5 years.
Seriously, our research highlights limitations in academic search engine indexing and querying. We call for an action from IEEE and ACM to better index rare terms. Meanwhile,  Google Scholar should provide a way to query the hyperlink graph embedded in academic documents.

Future work is required to fully understand the rickrolling phenomenon. For instance, preliminary inquiry suggests that it is used as fake persona homepages\footnote{As Github user profile link:
\url{https://github.com/search?p=3&q=dQw4w9WgXcQ&type=Users}}. And we leave for future work to survey the academic literature on ``Dance Dance Authentication''\footnote{\url{https://scholar.google.se/scholar?q=\%22VgC4b9K-gYU\%22}}.

\newpage

\bibliographystyle{IEEEtran}
\bibliography{rickroll.bib}

\end{document}